\documentclass{article}

\usepackage{graphicx}

\usepackage[preprint]{neurips_2024}

\usepackage[utf8]{inputenc} 
\usepackage[T1]{fontenc}    
\usepackage{hyperref}       
\usepackage{url}            
\usepackage{booktabs}       
\usepackage{amsfonts}       
\usepackage{nicefrac}       
\usepackage{microtype}      
\usepackage{xcolor}         
\usepackage[numbers]{natbib}
\usepackage{ulem}
\usepackage{enumitem}

\title{A theory of neural emulators}

\author{%
  Catalin C. Mitelut\\
  Forum Basiliense, University of Basel \\
  Foresight Institute \\
  \texttt{mitelutco@gmail.com} \\
}

\begin{document}

\maketitle

\begin{abstract}
A central goal in neuroscience is to provide \textit{explanations} for how animal nervous systems can generate actions and cognitive states such as consciousness while artificial intelligence (AI) and machine learning (ML) seek to provide models that are increasingly better at \textit{prediction}. Despite many decades of research we have made limited progress on providing neuroscience explanations yet there is an increased use of AI and ML methods in neuroscience for prediction of behavior and even cognitive states. Here we propose emulator theory (ET) and neural \textit{emulators} as circuit- and scale-independent predictive models of biological brain activity and \textit{emulator theory} (ET) as an alternative research paradigm in neuroscience. ET proposes that predictive models trained solely on neural dynamics and behaviors can generate functionally indistinguishable systems from their sources. That is, \textit{compared to the biological organisms which they model, emulators may achieve indistinguishable behavior and cognitive states - including consciousness - without any mechanistic explanations}. We posit ET via several conjectures, discuss the nature of endogenous and exogenous activation of neural circuits, and discuss neural causality of phenomenal states. ET provides the conceptual and empirical framework for prediction-based models of neural dynamics and behavior without explicit representations of idiosyncratically evolved nervous systems.

\end{abstract}

\section{Introduction}

A central goal of neuroscience research is to provide \textit{explanatory models} \cite{marr1982vision, churchland1992computational} underlying brain function from molecules and genetics to memory encoding, decision making and cognition \cite{shen2022emergence}. After more than a century of extensive research in anatomy and physiology we understand anatomical connectivity better - but have made limited progress in understanding function, especially at the whole-organism scale, including how multi-scale dynamical interactions within the central nervous system (CNS) give rise to our behaviors and phenomenal experiences \cite{roland2023far}. 
Our limited progress may be due to avoiding difficult paradigms involving behavior \cite{krakauer2017neuroscience, humphries2017neuroscience}, being disconnected from psychology \cite{beste2021disconnected}, or lacking large datasets and powerful models \cite{markram2013seven}. Recent controversies including possibly pursuing the wrong approach in Alzheimer's research for decades \cite{piller2022blots} raise the question whether seeking to provide completely mechanistic models of brain function will ever succeed at the whole organisms level, help solve complex diseases or explain capacities such as consciousness.

In parallel, over the past several decades deep machine learning (deep ML) and artificial intelligence (AI) methods relying on black-box neural networks (NNs) have created increasingly powerful \textit{predictive models} that can achieve superhuman game performance \cite{silver2016mastering}, and improved language translation \cite{popel2020transforming} and language generation \cite{radford2019language}, among many other feats \cite{lecun2015deep, alzubaidi2021review}. While the value of explanatory vs. predictive models has been debated in statistics for decades \cite{Breiman2001}, it is also being debated in newer fields, e.g. neuroscience \cite{bowers2023deep, lin2023scientific} and psychology \cite{yarkoni2017choosing}. Some neuroscience researchers have begun implementing NNs for \textit{predictive} modeling of neural systems \cite{richards2019deep, kietzmann2019deep} showing they are as good or better than explanatory or mechanistic models \cite{cichy2019deep} including for decoding of spatial location \cite{tampuu2019efficient}, latent dynamics \cite{zhou2020learning} or even inferring single trial population dynamics \cite{pandarinath2018inferring}. 

While it is still uncertain how much predictive models will contribute to our understanding of the brain - large-scale neural datasets are increasingly common and AI and ML methods are likely required to understand them. In fact, the extraordinary predictive power of Large-Language-Models (LLMs) \cite{Openai2023} in the last few years provides some evidence that prediction-only models offer some utility for scientific and knowledge dissemination. Additionally, given their success some have even speculated whether future LLMs could experience conscious states \cite{butlin2023consciousness}. These and other debates have contributed to an already existing interdisciplinary research field on "machine" or "artificial" consciousness \cite{GAMEZ2008}. Deciding on the usefulness of modern AI in neuroscience as well as their conceptual properties is challenging as we need to resolve conceptual, computational and experimental problems involving questions from AI, philosophy and neuroscience.



\uline{Scope of work: emulators as a novel type and use-case for predictive models}. Here we provide a novel framework for how \textit{predictive} models can contribute to neuroscience and whether such models could generate internal states such as consciousness. We propose the concept of \textit{neural emulators} (or "emulators" for short): predictive (joint) models of animal behavior and whole-organism neural dynamics generated from different spatio-temporal scaled data (Fig 1).  In their simplest formulation, emulators predict behavior and neural dynamics based on historical neural and behavioral states and can be used to study the informational content and structure of neural activity in specific areas, across areas and with respect to behavior. In their most complete form, emulators learn to model the (nearly) complete causal structure of neural-behavior interactions and can generate outputs that are indistinguishable from the behaviors and phenomenal states of the organisms they model. 

\begin{figure}

  \centering
  \fbox{\rule[-.25cm]{0cm}{0cm}   \rule[-.25cm]{0cm}{0cm}
  \includegraphics[width=.55\linewidth]{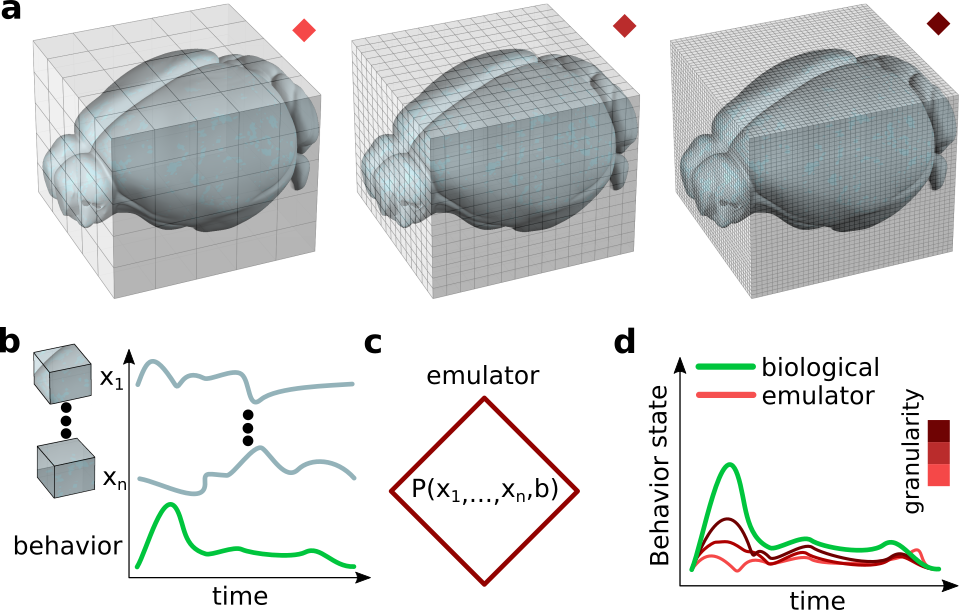}}

  \caption{\textbf{Circuit- and scale-agnostic neural emulators}. \textbf{(a)}. Recording from a rodent brain (light-blue) based on parcelation (cubes) of increasing granularity (red-hue diamonds). \textbf{(b)}. Neural time series from parcellation in (a) (light blue) and behavior time series (green). \textbf{(c)}. Emulators learn joint probability of time-series in (b). \textbf{(d)}. Behavior output of emulators is increasingly similar to  biological behavior as parcellation granularity increases.}  

\end{figure}

\uline{Contribution of work: emulators and emulator theory}. We define emulators and propose \textit{emulator theory} (ET) as a theoretical framework for measuring and replicating the capacities of biological organisms solely via \textit{predictive} models that do not require explicit mechanisms of how neural (sub)systems interact. ET argues that neural dynamics and causality can be simulated or artificially generated and proposes that emulators can capture all capacities of biological brains including possibly conscious experiences. The scope of our work and contributions are as follows:
 
\begin{enumerate}[itemsep=-2pt] 
    \item provide a definition of neural emulators, describe the axioms and conjectures of emulator theory (ET) and provide a description for constructing neural emulators from neurophysiology datasets (Section \ref{sec:neural_emulators} and Appendix \ref{sec:ideal_emulators}).
    
    \item propose a theory of neural causality that explain how models trained solely on prediction - such as emulators - can generate indistinguishable behavior and first person phenomenal states of biological organisms (Section \ref{sec:path_independence_causality}) and Appendix \ref{sec:physics_of_causality}, 
    and \ref{sec:MSD_model_indistinguish}).
\end{enumerate}

\uline{Relation to previous work}. Our work is related to several research paradigms including: machine consciousness, phenomenology, philosophy and physics of causality, and deep ML methods for modeling time-series data. We discuss our work in relation to previous research at length in Appendix \ref{sec:previous_work}. Briefly, in comparison to previous work, we propose a novel interpretation for predictive models of neural activity that is independent of architecture or mechanistic explanations of brain function.

\section{A framework for neural emulators}
\label{sec:neural_emulators}
Here we define the central components of emulator theory (ET) - a proposal for how and why artificial (i.e. computational) systems trained to predict the behavioral and cognitive states of animals can become increasingly accurate copies, or "emulators", of such organisms.  
In particular, ET proposes that models trained on sufficiently large datasets solely to predict behavior and phenomenal states have the capacity to generate such states - even without access to explanatory mechanisms.\footnote{We henceforth omit the terms "sufficiently large" datasets and discuss further in Appendix \ref{sec:terminology}}.  
ET proposes this can occur because both behaviors and phenomenal states are causal-path independent: i.e. the neural states supporting behavior and consciousness do not depend on a specific neuronal pathway - only on specific neuronal dynamics.

Below we provide a framing of ET relative to computational neuroscience approaches, describe the causal-path independence conjecture and provide a practical approach to constructing emulators. 





\subsection{Common assumptions in explanatory models of behavior and neural dynamics}

A central goal for computational and behavior neuroscience research is to identify \textit{explanatory} mechanisms for how the nervous systems of animals can generate behaviors and cognition. We generalize these approaches as seeking functional descriptions of (i) behaviors and (ii) neural dynamics.\footnote{We note that reportability of conscious states and presence of conscious states is a topic in philosophy of mind and psychology broadly under several topics including "phenomenal" versus "access consciousness" \cite{naccache2018}. Here we focused on "reportable" only for the purpose of having a simplified training label for our models though this is not central to our argument and other labels would suffice. }:
\vspace{-0.5em}

\begin{equation}
F_t^b(x_{i,t}) = \mathrm{Observed\ behavior}
\end{equation}
\begin{equation}
    F_t^c(x_{i,t}) = \mathrm{(Reportable)\ state\ of\ consciousness}
\end{equation}
with:
\begin{equation}
    x_{i,t} = Q(x_{i,t_{-1}}, s_{t_{-1}})
\end{equation}

where $F_t^b$ and $F_t^c$ describe how behavior and conscious states, respectively, are generated from $x_{i,t}$ which are measurements of neural component $i$ (most often the spiking of a single neuron)  at time $t$, and 
$Q$ describes the evolution of the neural components from endogenous (i.e. $x_{i,t_{-1}}$) and exogenous inputs (or stimuli) $s_{t_{-1}}$. Thus, for example, we want to show how neural activity in motor cortex \cite{churchland2012neural} and supporting areas \cite{SVOBODA201833} generate observable body movements, even at the single trial level \cite{pandarinath2018inferring}, by describing how neural states (e.g. spiking of populations of neurons) generate observables such as location of the arm or hand of an animal or the location of a rodent in an environment. 

This framework is common in computational and theoretical neuroscience (though it is not the only one) and makes several assumptions:

\begin{enumerate}
    \item $Q$, $F_t^b$ and $F_t^c$ have a closed form expression that we can eventually discover.

    \item $Q$, $F_t^b$ and $F_t^c$ will describe necessary causal mechanisms without which neither behaviors nor cognitive states can occur.

    \item Identifying $Q$, $F_t^b$ and $F_t^c$ requires single neuron or lower spatial scale neural data.
    
\end{enumerate}

\subsection{Emulators are mechanism- and spatial scale-independent predictive models of behavior and neural dynamics}

ET proposes to eliminate all three assumptions from building useful models of the brain. In particular, ET propose that both behaviors and cognitive states can be accurately predicted by models that do not represent or instantiate $Q$, $F_t^b$ or $F_t^c$ and without an explicit spatio-temporal scale of neural activity. Thus ET proposes that at a specific neural data granularity $g$ we can build emulator $E$ to predict behaviors and conscious states:

\begin{equation}
    E_t^b(w_{j,t,g}) = \mathrm{Observed\ behavior}
\end{equation}
\begin{equation}
    E_t^c(w_{j,t,g}) = \mathrm{(Reportable)\ state\ of\ consciousness}
\end{equation}
\begin{equation}
    w_{j,t,g} = R(w_{j,t_{-1},g}, s_{t_{-1}})
\end{equation}

where $w_{j,t,g}$ is the state of parcel $j$ of a system sampled at granularity $g$ and at time $t$ that evolves dynamically and $E_t^b$, $E_t^c$ and $R$ are transformation learned from "big" data. Thus, for example, emulators can be described simply by nested (black-box) neural-network (NN) models:

\begin{equation}
    \mathrm{NN}_2(\mathrm{NN}_1(w_{j,t,g})) = \mathrm{Observed\ behavior}
\end{equation}
\begin{equation}
    \mathrm{NN_3}(\mathrm{NN_1}(w_{j,t,g})) = \mathrm{(Reportable)\ state\ of\ consciousness}
\end{equation}

where NN$_{1}$, NN$_{2}$ and NN$_{3}$ stand for $R$, ${E_t^b}$, and ${E_t^c}$, respectively. 

In this framework, emulators are predictive models that are imperfect - i.e. have imperfect prediction - but can improve their accuracy with increased granularity of neural data and size of training datasets. This is similar to many, if not most, ML approaches in data modeling. However, there are at least two major differences. First, in the large dataset and high granularity limit ET claims that emulators can generate phenomenal states even without an explanation of how phenomenal states are caused. Second, and more simply, ET claims that despite the extreme complexity of the brain,  datasets generated in practical experiments (e.g. finite-time laboratory neuroscience experiments) are sufficient for building accurate emulators.  In the remainder of this section we address these two issues. 

\subsection{Emulators (must) model causality of phenomenal states}

Emulators are more than (black-box) models that seek to predict behaviors or conscious states - they are models that generate such outputs \textit{while modeling the neural dynamics causing the outputs}. In this sense, unlike general NNs fit to data - emulators implicitly capture \textit{the idiosyncratically evolved} neural causal pathways by jointly modeling neural dynamics and observables. 

But shouldn't our models (predictive or mechanistic ones) first explain in mechanistic terms the causal pathways for generating behaviors or conscious experience before being able to generate them? In our view, this requirement may be too conservative as it suggests \textit{a priori} that (human) understanding of neural causality must precede generation of neural causality. In our view, the question of \textit{how we can capture} the causal interactions of a neural system to recreate its behavior and internal states - is an empirical question, and one that ET in part seeks to address. 

\subsection{Exogenous generation of behavior and phenomenal states is already possible}

In fact, our mechanistic understanding of neural causality already lags behind the ability to artificially generate behaviors or even conscious states in biological organisms. In particular, we have known for almost a century how to artificially generate conscious states via direct, i.e. exogenous, electrical stimulation (DES). For example, in awake human neuro-surgery, activations of neural tissue by direct (i.e. exogenous) electrical stimulation can lead to individuals experiencing specific conscious content: emotions and novel visual imagery \cite{lai2020acute}, speech generation \cite{collee2023localization}, or even disrupting short term memory of the task itself \cite{ng2021disrupting}. An even more remarkable finding in the neuroscience of agency is that even self-control states such as experiencing a desire or "will to act" can be elicited by exogenous electrical stimulation \cite{fried1991functional}. In mice, using light optogenetic activation \cite{deisseroth2015optogenetics} we can identify and then artificially activate visual system neurons to generate perceptions leading to behavior \cite{marshel2019cortical, CARRILLOREID2019447}. We can even "tag" memory engrams representing positive moods and reactivate them at a later time to elicit mood-change like behaviors in mice \cite{ramirez2015activating}. 

Even though it is very likely that both DES and optogenetic activations engage specific - rather than random - neural circuits and states, these are nonetheless remarkable experimental findings. That is, \textit{there is no obvious requirement that evolution required that organism-level behaviors or phenomenal states can be generated simply by exogenous input} - rather, the opposite seemed more possible given the extraordinary complexity of the brain. Namely, that highly specific neurons and circuits must be engaged in a particular fashion to generate behaviors and phenomenal states. In our view, the existence of coarse-grained exogenous generation of phenomenal states (and behavior) is a largely under-explored research approach to understanding behavior  and consciousness causality. More specific to ET, these studies suggest that - at the very least - we may not require precise mechanistic models of neural dynamics all the way to the synapse (or lower level) to generate models that exhibit behavior and phenomenal states. 


\subsection{Causal path independence conjecture}

More importantly, these findings suggest that neural states that support the generation of actions or conscious states may be reacheable by multiple paths. We formulate this into a central conjecture of ET:
\begin{enumerate}[label={C\arabic*:}]

    \item \textbf{Path independent neural causality conjecture (PINC)}. Behaviors and conscious states do not dependent on endogenous (i.e. within system) causality  and they can be generated (or reached) by completely exogenous pathways.
    
\end{enumerate}

Here, "exogenous" means a perturbation that is external to the brain such as a patch clamp or optogenetic excitation. In practical terms, $C1$ proposes that sequences of neural activation $x_{i,t_{n}}$ ($n$ time steps) that are sufficient for conscious experience or behaviors - can be generated (i.e. caused) by exogenous sequential activation of such neural states. 
Thus, $C1$ implies that any mechanism required for consciousness, for example the "ignition and broadcast within a neuronal global workspace" (GWT), "local recurrent or re-entrant cortical processing" (recurrence), or  "multiple ... representations ... available to a central system" (multiple drafts)\cite{seth2022theories} - can be activated by exogenous pathways as \textit{there is no dependence of conscious states emerging from specific causal paths, but only on dynamical activation}. 

We view PINC as central to understanding how causality can be emulated in a artificial systems (such as computational models or emulators) which can in turn generate behaviors and conscious states in biological and even artificial organisms. We discuss PINC in more detail in Section \ref{sec:path_independence_causality}. 



\begin{figure}
  \centering
  \fbox{\rule[-.25cm]{0cm}{0cm}   \rule[-.25cm]{0cm}{0cm}
  \includegraphics[width=.6\linewidth]{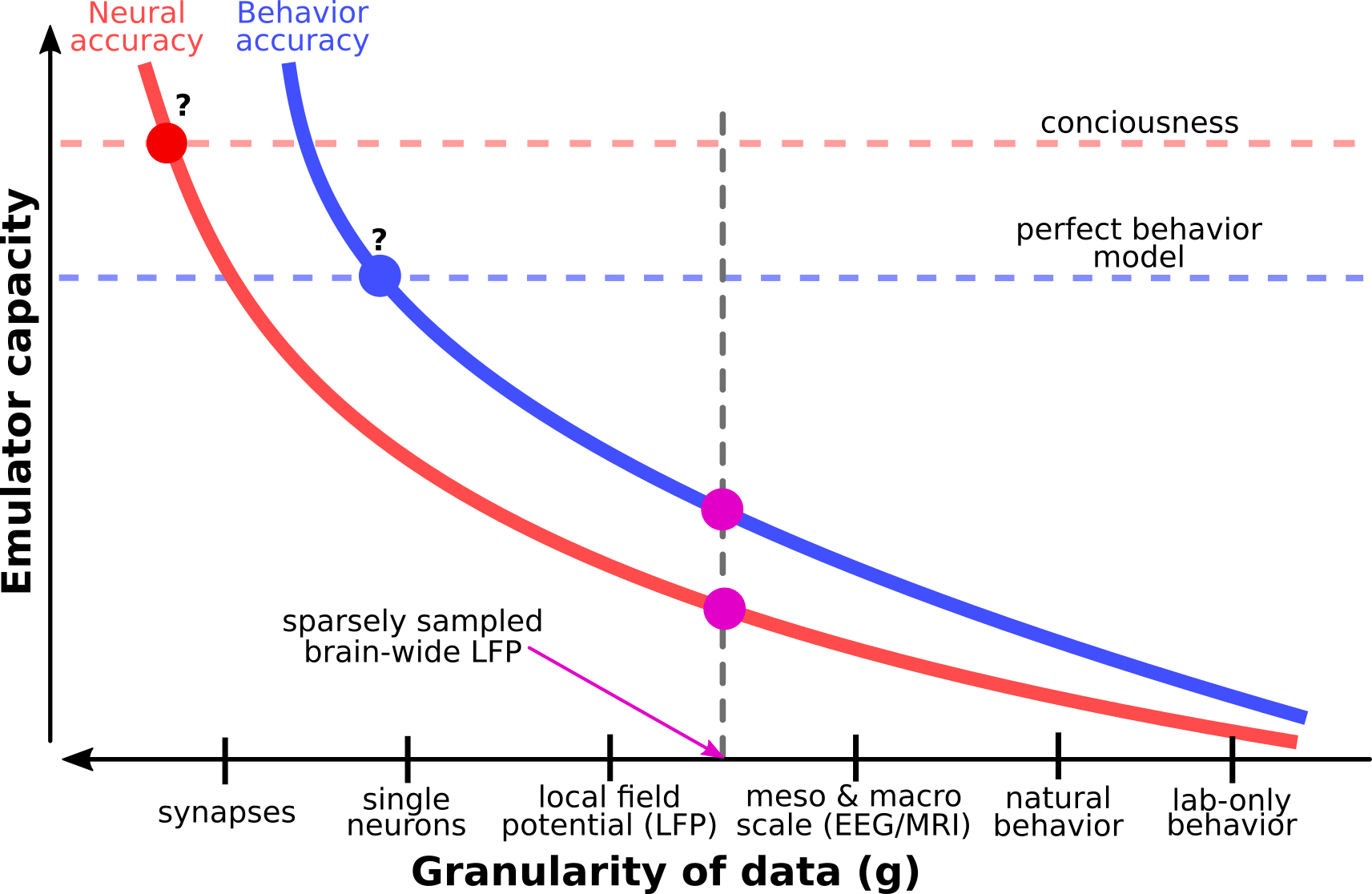}}
 
  \caption{\textbf{Granularity-based neural emulators.} Relationship between the capacity (or accuracy) of a behavior (blue curve) and neural (red curve) emulator vs the granularity of the neural data used for training with hypothesized requirements for perfect behavior models (dashed blue line) and conscious states (dashed red line). Proposed experiment of recording sparsely sampled brain-wide LFP (magenta arrow) and putative emulator capacity from such datasets (magenta dots).}
  \label{fig:granularity_emulators}
\end{figure}

\subsection{Scalable emulators: a research paradigm}

We now return to the question of whether we can collect sufficient datasets for training mechanism-agnostic emulators. Central to this question is how to build emulators from empirical data. We find it instrutive to start by defining an \textit{ideal emulator} as an abstract tool to help frame the practical emulators. An ideal emulator is a model that is generated from very low (or arbitrarily small) spatial-scale granularity parcels and very large (or arbitrarily large) datasets. We point out that ideal emulators seek to model the behavior of biological system \textit{but only to the level required to predict the behaviors and phenomenal states} and may not need to model synapse, or protein-level interactions in nervous systems. In this sense, ideal emulators model (only) the sufficient conditions for the generation of such observables. We discuss ideal emulators in more detail in Appendix \ref{sec:ideal_emulators}.

In contrast to ideal emulators we propose building \textit{scalable emulators} (or simply "emulators").  Scalable emulators can be created by training predictive models from (practical size) neural dynamics datasets of varying spatial and temporal resolution or "granularity" (Fig \ref{fig:granularity_emulators}). Emulators can thus be generated using neural activity recorded at different granularity scales such as: single neuron spikes, local-field-potential (LFP), or functional magnetic resonance imaging (fMRI). Emulators will thus be trained on (simultaneously recorded) neural dynamics and behaviors of biological organisms - to predict behavior and neural dynamics.  Does this mean that any type of data, e.g. functional magnetic resonance imaging (fMRI) is sufficient to generate perfect or nearly perfect emulators? 

Not necessarily.  We define the \textit{capacity} of an emulator simply as the accuracy of predicted behaviors and neural dynamics relative to the source (e.g. on test or hold out data). ET proposes that - much like any other ML model - emulator capacity or accuracy will increase with dataset size (e.g. number of time points and number of neural areas) - and also with increased granularity of the recordings (e.g. LFP models should do better than fMRI ones). While we do not expect that current types of fMRI data alone is sufficient to achieve perfect emulators, ET implies that given sufficiently large fMRI datasets there is no theoretical prohibitions on this. 

We close by providing a comparison between the dataset sizes used to train LLMs and what is feasibly possible to generate using relatively modest empirical paradigms in rodent neuroethology (Table \ref{tab:mytable}). Considering each individual neural parcel as a single token, and 1000Hz sampling rate, we can in principle generate datasets of similar size used to train GPT-4 (estimated) with just over a dozen rodents recording continuously for 24 months. 

\begin{table}[h]
    \centering
    \begin{tabular}{|c|c|c|c|c|}
        \hline
        Type of model & Recording time & No. of neural parcels & No of params & No. of training tokens\\
        \hline
         Llama 2 & n/a & n/a  & 7-70 billion & 5 trillion \\
        \hline
        GPT 4 & n/a & n/a & 1.76 trillion & O(100) trillion (est) \\
        \hline
        1 Rodent & 30 days & 32  & n/a & O(100) billion \\
        \hline
        15 Rodents & 2 years & 128 & n/a & O(100) trillion \\
        \hline
    \end{tabular}
    \caption{Comparison between training datasets for LLMs and emulators.}
    \label{tab:mytable}
\end{table}






\subsection{ET: Summary and Conclusion}
We proposed emulators as scale-dependent predictive models of neural dynamics and behaviors of biological organisms. We offered a conceptual theoretical framework, i.e. ET, which suggests that sufficiently accurate models, i.e. emulators, may generate artificial systems with the same capacities of the biological organisms they model. We argued this occurs because accurate emulators must instantiate models of neural dynamics that are similar to the biological nervous systems they model - essentially emulating neural causality of such biological nervous systems. We proposed a practical research framework - scalable emulators that focuses on modeling of simultaneously recorded neural and behavior data to generate increasingly accurate predictive models as a function of the granularity or spatio-temporal scale of the recordings.




One advantage of ET framework for modeling nervous systems is that we can \textit{balance the reductionist desire to decompose all neural states into the smallest components with principled pragmatic-driven goals}. Thus, we can remain agnostic on both mechanisms as well as spatio-temporal scale of recordings required to adequately model biological neural networks. In contrast to mechanistic models sought in neuroscience, ET states we don't need to understand the role of all neural components across all scales \textit{if} we can identify scales that enable us to generate sufficiently accurate models. Because there are redundancies in biological neural networks (e.g. not all neurons or areas are required for specific capacities), we may directly pursue model accuracy as a primary goal rather than mechanistic explanation of all components. 




In the next section we expand our discussion of the central ET conjecture - $C1$ - to provide a conceptual background for why neural states are causal-path independent and why this further supports the exogenous emulation of causality proposed by ET. We discuss additional topics on the philosophy of causality in the context of ET in Appendix \ref{sec:physics_of_causality}. 

\section{A theory of path-independent neural causality}\label{sec:path_independence_causality}

The central conjecture of ET, i.e. the path-independent-neural-causality (PINC) proposes that we can generate behaviors and conscious states using exogenously pathways as opposed to endogenous (or "natural") ones. We reviewed DES and optogenetic studies in humans and mice, respectively, and argued this provides some evidence for this. However, in our view, to fully establish emulators as exogenous models of causality we must do more than merely argue that some neural states can be generated exogenously. In particular, we must show that emulators can learn to activate biological - or artificial - circuits using models that capture and replicate causality. 

In its simplest form PINC highlights limits in empirical measurement and "measurability" of biological dynamical systems that generate phenomenal states. For example, if we could somehow activate any neural state ("down to the synapses") by external means - it may not be possible to measure a difference between such neural states arising naturally and exogenously. This is because the fundamental quantities we measure are behavioral or internal state "readouts" - and if those are nearly identical then we cannot in principle differentiate between their causes. 

A second, related, challenge that PINC raises is about "identicality": i.e. we may have to rethink what makes for a perfect or "ideal" model of a neural system or biological organisms. The contribution of chaos and noise to neural dynamics has been studied for decades in neuroscience even at the single neuron level \cite{Mainen1995} and in drift diffusion models of decision making where the precise timing of a decision is generally determined by noise \cite{Ratcliff2001, Ratcliff2008}. These studies (many non-mechanistic) propose broad boundaries on what constitutes a sufficiently accurate model of an observed neural system - or its behavioral output. Thus, PINC raises the question of what constitutes a perfect model - even one built from mechanistic knowledge - of a biological organism?

Below we build our intuition for PINC using various levels of abstract thought experiments attached to hypotheses. We start with the \textit{input source indistinguishability} (ISI) hypothesis which states that from a first person perspective it is impossible to distinguish whether the behavior output or conscious state of an organism was generated by endogenous or exogenous pathways. Second, the \textit{path-independent causality} (PIC) hypothesis restates $C1$ (see Section \ref{sec:neural_emulators}) in the framework of causal models and highlights the lack of causal-path requirements for neural system dynamics. Lastly, the \textit{model system divergence} (MSD) hypothesis proposes that because of noise and chaotic dynamics - there can be no \textit{a priori} thresholds for distinguishing between the output of a dynamical system driven by a (sufficiently accurate) emulator and a system driven by ground-truth mechanistic model. 

\subsection{ISI: all neuronal states can be reached exogenously}
\begin{figure}
  \centering
  \fbox{\rule[-.25cm]{-0.25cm}{-1.5cm}   \rule[-.0cm]{0cm}{0cm}
  \includegraphics[width=.98\linewidth]{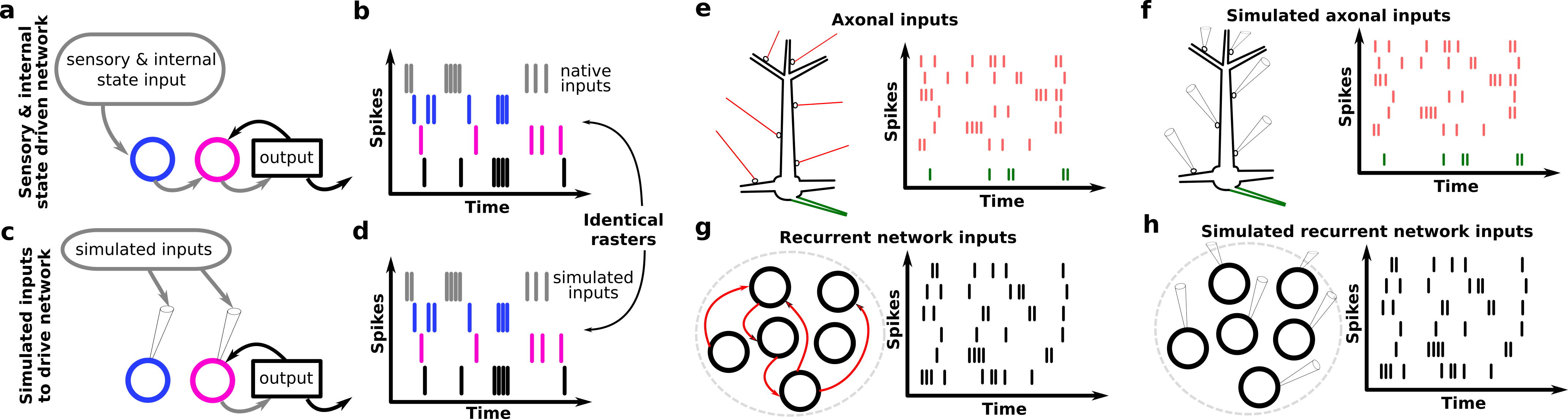}}
 
  \caption{\textbf{Input source indistinguishability (ISI)}. \textbf{(a)} A two neuron local network receiving "native" sensory and internal state input and generating downstream output. \textbf{(b)} Spike rasters for network in (a). \textbf{(c)} Same as (a) but inputs to network are simulated by an external system. \textbf{(d)} Same as (a) but for simulated system as in (c) results in identical spike rasters. \textbf{(e)} Left: single neuron receiving axonal inputs (red) from various sources and outputting a spike pattern (green); Right: proposed spike rasters for (e). \textbf{(f)} Left: same as (e) but axonal inputs are simulated by external system; Right: identical raster to (e) achieved by simulated system in (f). \textbf{(g)} Left: a self-driving recurrent neural network; Right: proposed spike rasters for (g). \textbf{(h)} Left: same as (g) but with recurrent inputs removed and simulated inputs; Right: identical spike raster as in (g) achieved by simulated system in (h).}  
  \label{fig:isi}

\end{figure}

We begin with a model of a 2-neuron circuit receiving input from sensory and internal state systems (i.e. "native" input) and generating downstream output (Fig \ref{fig:isi}a). From an external perspective (i.e. the general perspective of an experimental neuroscientist) this system can be completely described by the spike trains of all the nodes (Fig \ref{fig:isi}b). Having access to such spike trains - we can imagine removing (or ablating) the native connections to this circuit and directly or "artificially" simulating inputs (e.g. using patch clamps) (Fig \ref{fig:isi}c). Such an artificially driven system would have identical\footnote{We assume determinism here, without loss of generality.} spike rasters with the original system (Fig \ref{fig:isi}d). More importantly, as our artificially driven system is functionally identical to our native one - \textit{from a first person perspective there is no experiment that we can do to determine whether the neural states are generated by native or artificial inputs}. 

This thought experiment can be modified \textit{ad infinitum}. For example, we can consider smaller size scales by modeling synapse activations (Fig \ref{fig:isi}e,f) and we can consider complex recurrent neural networks by increasing the complexity of the connectivity (Fig \ref{fig:isi}g,h). In all cases, if we can adequately model the neural dynamics and artificially generate them - from a first person behavior it is not possible to determine how causality was achieved.

Does this argument extend to arbitrary scales, e.g. the whole brain - and what are the implications for cognition or consciousness? With respect to viability, single neuron activations is increasingly achievable \cite{tong2023single}. At the phenomenal level, ISI hypothesizes that all neural states including those underpinning conscious states can can be achieved by artificial activation of neurons even in complex recurrent networks. More specifically, ISI proposes that irrespective of the necessary neural states for consciousness - such states can be artificially generated with a sufficiently powerful model of neural activity and precise activation (see discussion above on DES). 



\subsection{PIC: all neural states are causal path agnostic}

Conscious states are thought to unfold on the scale of hundreds of milliseconds to seconds \cite{northoff2020neural, kent2019duration} and most neuroscience theories of consciousness require complex interactions across many neural systems where temporal dynamics matter \cite{seth2022theories}. Given that we may not be able to distinguish the source of neural inputs - does this mean that dynamics (e.g. order of activation or recurrence) do not matter for complex states such as cognition or consciousness? 

The path-independent causality conjecture (PIC) builds on ISI and proposes that while neural dynamics may matter for the generation of conscious states, the \textit{causal paths do not}. That is, from a first person perspective - it is impossible to distinguish the causal pathway by which the necessary and sufficient neuronal states for consciousness are achieved. Thus, at best, only temporal dynamics matter and that causal pathways can be simulated with sufficient manipulation. How is this possible? 

\begin{figure}
  \centering
  \fbox{\rule[-.25cm]{0cm}{0cm}   \rule[-.25cm]{0cm}{0cm}
  \includegraphics[width=.75\linewidth]{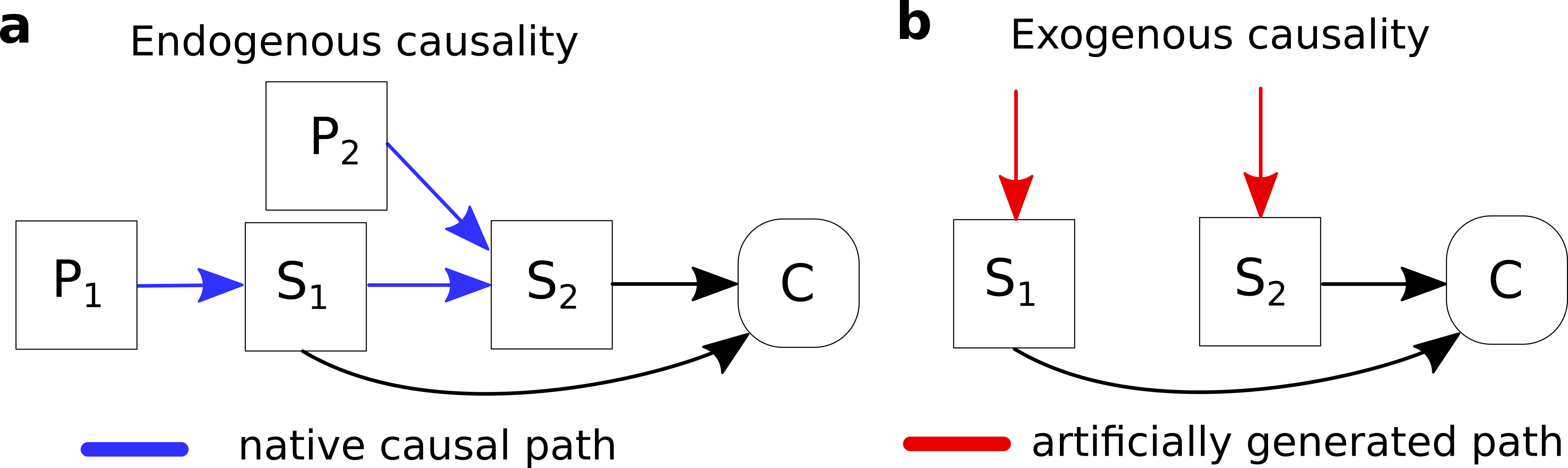}}
 
  \caption{\textbf{Path-independent causality conjecture (PIC)}. \textbf{(a)} Causal graph of two necessary and sufficient biophysical neural states, S$_1$ and S$_2$, required for phenomenal conscious state C. In the native state, neural states S$_1$ and S$_2$ are caused by endogenous prior states P$_1$ and P$_2$ (e.g. sensory and internal state processing). \textbf{(b)} Same as (a) but  neural states S$_1$ and S$_2$ are generated via exogenous (e.g. artificially generated) inputs resulting  in the same phenomenal cognitive state C irrespective of the causal path of activation of necessary and sufficient biophysical neural states.}
  \label{fig:pic}

\end{figure}

To show this we study a causal model of conscious state generation (Fig \ref{fig:pic}). In particular, we study the causal generation of sufficient neural states S$_1$ and S$_2$ for the generation of conscious state C. In this framework the \textit{direct causes} (i.e. parents) of C are S$_1$ and S$_2$ which are in turn caused by their own parents (P$_1$ and P$_2$; Fig \ref{fig:pic}a). Thus, given S$_1$ and S$_2$, C is independent of the parents:

\vspace{-8pt}
\begin{equation}
    C \perp \!\!\! \perp (P_1, P_2) \mid (S_1, S_2)
\end{equation}

In this causal framework, because C is caused solely by S$_1$ and S$_2$, the causes of by S$_1$ and S$_2$ can be arbitrarily set or determined (Fig \ref{fig:pic}b).\footnote{For clarity, objections that "conscious" states require many other interacting parts, including states such as P$_1$ and P$_2$ to be activated, we simply add more exogenous causal paths to represent their activations as well.} Thus, if we replace endogenous input (i.e. parents P$_1$ and P$_2$) with exogenous input - conscious state $C$ is still achieved. \footnote{We note this description generalizes to all possible neural states from simple two neuron systems to large recursively activated networks. For example, in HOT S$_1$ can represent lower order representations and S$_2$ higher order, or in GWT S$_1$ can represent globalization of information and S$_2$ attention being brought to the states.}  






\subsection{MSD: emulator outputs may be indistinguishable from mechanistic models and biological organism outputs}

Our last hypothesis - MSD - states that there are no \textit{a priori} methods for distinguishing between sufficiently accurate predictive models (i.e. emulators) and models built from a complete mechanistic theory of neurvous system function. Fundamentally, this hypothesis suggests that \textit{we must define empirical tests and thresholds} for what constitutes a sufficiently accurate model of a biological neural system. We discuss this in more detail in Appendix \ref{sec:MSD_model_indistinguish}.

\section{Limitations}

Our proposal has some limitations. In particular, ET does not describe the content of conscious states or the necessary and sufficient conditions for generating them. ET primarily states that such states can be achieved via artificial activations - or ultimately within completely artificial systems that model biological neural networks in sufficient detail. Additionally, ET proposes that experimentally viable emulators are possible but it is an empirical question that needs to be investigated. Similarly, we did not discuss the size of the datasets required but also the neural-area specificity of our datasets. We envision that neural data from some areas such as frontal and sensory systems (known to be involved in conscious state and behavior generation) may provide more information for our models. In this sense, building emulators in the near-term (i.e. before technologies to record from the whole brain are available) should benefit from neuroanatomical targeting.


Lastly, we did not discuss neural architectures required for generation of conscious states. ET is agnostic to the role of architectures (or even anatomical modularity) in generating behavior and conscious states. In particular, ET does not make any commitments to the requirement of specific neural circuits - but only indirectly supports whatever artificial NN architectures are required to model of causality. However, given the universal approximation theorem, given sufficiently expressive NNs and large enough datasets - ET suggests architectures do not matter. In this context, ET may be interpreted to suggest that biological organisms developed specific neural architectures due to biophysical and energy constraints not potentially required for artificial or synthetic NNs. 

\label{sec:limitations}

\section{Concluding remarks}\label{sec:conclusion}
Here we proposed emulators as predictive models of organism behavior and cognitive states trained solely on historical neural dynamics and behavioral states. We offered ET as a framework for understanding how emulators can be built, what they can achieve and suggested benefits and limitations. ET also points to the need for developing increasingly nuanced empirical tests to determine whether models of complex systems exhibit the capacities of the originals including internal phenomenal experiences. In our view, once models of biological neural networks achieve high precision - it is not only possible that they experience conscious states - but  likely that they do so.  

\newpage
{\small
\bibliography{neurips_2024}

\begin{thebibliography}{10}

\bibitem{marr1982vision}
David Marr.
\newblock {\em Vision: A Computational Investigation into the Human
  Representation and Processing of Visual Information}.
\newblock The MIT Press, Cambridge, MA, 1982.
\newblock Foreword by Shimon Ullman, Afterword by Tomaso A. Poggio.

\bibitem{churchland1992computational}
Patricia~Smith Churchland and Terrence~J Sejnowski.
\newblock {\em The computational brain}.
\newblock MIT press, Cambridge, 1992.

\bibitem{shen2022emergence}
Y.~Shen, A.~Luchetti, G.~Fernandes, et~al.
\newblock The emergence of molecular systems neuroscience.
\newblock {\em Molecular Brain}, 15:7, 2022.

\bibitem{roland2023far}
Per~E. Roland.
\newblock How far neuroscience is from understanding brains.
\newblock {\em Frontiers in Systems Neuroscience}, 17:1147896, 2023.

\bibitem{krakauer2017neuroscience}
JW~Krakauer, AA~Ghazanfar, A~Gomez-Marin, MA~MacIver, and D~Poeppel.
\newblock Neuroscience needs behavior: Correcting a reductionist bias.
\newblock {\em Neuron}, 93(3):480--490, Feb 2017.

\bibitem{humphries2017neuroscience}
Mark Humphries.
\newblock Is neuroscience doomed to go round in circles? on the disturbing lack
  of direction in neuroscience.
\newblock {\em Medium}, May 2017.

\bibitem{beste2021disconnected}
Christian Beste.
\newblock Disconnected psychology and neuroscience—implications for
  scientific progress, replicability and the role of publishing.
\newblock {\em Communications Biology}, 4:1099, 2021.

\bibitem{markram2013seven}
Henry Markram.
\newblock Seven challenges for neuroscience.
\newblock {\em Functional Neurology}, 28(3):145--151, Jul-Sep 2013.

\bibitem{piller2022blots}
Charles Piller.
\newblock Blots on a field?
\newblock {\em Science}, 377(6604):358--363, Jul 2022.

\bibitem{silver2016mastering}
David Silver, Aja Huang, Chris~J. Maddison, Arthur Guez, Laurent Sifre, George
  van~den Driessche, Julian Schrittwieser, Ioannis Antonoglou, Veda
  Panneershelvam, Marc Lanctot, Sander Dieleman, Dominik Grewe, John Nham, Nal
  Kalchbrenner, Ilya Sutskever, Timothy Lillicrap, Madeleine Leach, Koray
  Kavukcuoglu, Thore Graepel, and Demis Hassabis.
\newblock Mastering the game of go with deep neural networks and tree search.
\newblock {\em Nature}, 529:484--489, Jan 2016.

\bibitem{popel2020transforming}
Martin Popel, Marketa Tomkova, Jakub Tomek, Łukasz Kaiser, Jakob Uszkoreit,
  Ondřej Bojar, and Zdeněk Žabokrtský.
\newblock Transforming machine translation: a deep learning system reaches news
  translation quality comparable to human professionals.
\newblock {\em Nature Communications}, 11, Sep 2020.

\bibitem{radford2019language}
Alec Radford, Jeff Wu, Rewon Child, David Luan, Dario Amodei, Ilya Sutskever,
  et~al.
\newblock Language models are unsupervised multitask learners.
\newblock {\em OpenAI blog}, 1(8):9, 2019.

\bibitem{lecun2015deep}
Yann LeCun, Yoshua Bengio, and Geoffrey Hinton.
\newblock Deep learning.
\newblock {\em Nature}, 521:436--444, 2015.

\bibitem{alzubaidi2021review}
Laith Alzubaidi, Jinglan Zhang, Amjad~J. Humaidi, Ayad Al-Dujaili, Ye~Duan,
  Omran Al-Shamma, J.~Santamaría, Mohammed~A. Fadhel, Muthana Al-Amidie, and
  Laith Farhan.
\newblock Review of deep learning: concepts, cnn architectures, challenges,
  applications, future directions.
\newblock {\em Journal of Big Data}, 8(53), 2021.

\bibitem{Breiman2001}
Leo Breiman.
\newblock {Statistical Modeling: The Two Cultures (with comments and a
  rejoinder by the author)}.
\newblock {\em Statistical Science}, 16(3):199 -- 231, 2001.

\bibitem{bowers2023deep}
Joshua~S Bowers, Gautam Malhotra, Marko Dujmović, et~al.
\newblock Deep problems with neural network models of human vision.
\newblock {\em Behavioral and Brain Sciences}, 46:e385, 2023.

\bibitem{lin2023scientific}
Hui Lin.
\newblock The scientific value of explanation and prediction.
\newblock {\em Behavioral and Brain Sciences}, 46:e399, 2023.

\bibitem{yarkoni2017choosing}
Tal Yarkoni and Jacob Westfall.
\newblock Choosing prediction over explanation in psychology: Lessons from
  machine learning.
\newblock {\em Perspectives on Psychological Science}, 12(6):1100--1122, 2017.

\bibitem{richards2019deep}
Blake~A. Richards, Timothy~P. Lillicrap, Philippe Beaudoin, Yoshua Bengio,
  Rafal Bogacz, Anders Christensen, Claudia Clopath, Rui~Ponte Costa, Archy
  de~Berker, Surya Ganguli, Colleen~J. Gillon, Danijar Hafner, Adam Kepecs,
  Nikolaus Kriegeskorte, Peter Latham, Geoffrey~W. Lindsay, Kenneth~D. Miller,
  Richard Naud, Christopher~C. Pack, Panayiota Poirazi, Pieter Roelfsema, Joao
  Sacramento, Andrew Saxe, Benjamin Scellier, Anna~Christina Schapiro, Walter
  Senn, Greg Wayne, Daniel Yamins, Friedemann Zenke, Joel Zylberberg, Dominic
  Therien, and Konrad~P. Kording.
\newblock A deep learning framework for neuroscience.
\newblock {\em Nat Neurosci}, 22(11):1761--1770, Nov 2019.

\bibitem{kietzmann2019deep}
Tim Kietzmann, Patrick McClure, and Nikolaus Kriegeskorte.
\newblock Deep neural networks in computational neuroscience.
\newblock {\em Oxford Research Encyclopedia of Neuroscience}, January 2019.

\bibitem{cichy2019deep}
Radoslaw~M. Cichy and Daniel Kaiser.
\newblock Deep neural networks as scientific models.
\newblock {\em Trends Cogn Sci}, 23(4):305--317, Apr 2019.

\bibitem{tampuu2019efficient}
A~Tampuu, T~Matiisen, HF~Ólafsdóttir, C~Barry, and R~Vicente.
\newblock Efficient neural decoding of self-location with a deep recurrent
  network.
\newblock {\em PLoS Comput Biol}, 15(2):e1006822, Feb 2019.

\bibitem{zhou2020learning}
Ding Zhou and Xue-Xin Wei.
\newblock Learning identifiable and interpretable latent models of
  high-dimensional neural activity using pi-vae.
\newblock In {\em Proceedings of the 34th International Conference on Neural
  Information Processing Systems (NeurIPS '20)}, pages 7234--7247, December
  2020.

\bibitem{pandarinath2018inferring}
Chethan Pandarinath, Daniel~J. O'Shea, Jasmine Collins, Rafal Jozefowicz,
  Sergey~D. Stavisky, Jonathan~C. Kao, Eric~M. Trautmann, Matthew~T. Kaufman,
  Stephen~I. Ryu, Leigh~R. Hochberg, Jaimie~M. Henderson, Krishna~V. Shenoy,
  L.~F. Abbott, and David Sussillo.
\newblock Inferring single-trial neural population dynamics using sequential
  auto-encoders.
\newblock {\em Nature Methods}, 15(9):805--815, 2018.

\bibitem{Openai2023}
OpenAI.
\newblock Gpt-4 technical report.
\newblock 2023.

\bibitem{butlin2023consciousness}
Patrick Butlin, Robert Long, Eric Elmoznino, Yoshua Bengio, Jonathan Birch,
  Axel Constant, George Deane, Stephen~M. Fleming, Chris Frith, Xu~Ji, Ryota
  Kanai, Colin Klein, Grace Lindsay, Matthias Michel, Liad Mudrik, Megan A.~K.
  Peters, Eric Schwitzgebel, Jonathan Simon, and Rufin VanRullen.
\newblock Consciousness in artificial intelligence: Insights from the science
  of consciousness.
\newblock {\em arXiv preprint arXiv:2308.08708}, 2023.

\bibitem{GAMEZ2008}
David Gamez.
\newblock Progress in machine consciousness.
\newblock {\em Consciousness and Cognition}, 17(3):887--910, 2008.

\bibitem{naccache2018}
Lionel Naccache.
\newblock Why and how access consciousness can account for phenomenal
  consciousness.
\newblock {\em Phil. Trans. R. Soc. B}, 373(20170357), 2018.

\bibitem{churchland2012neural}
MM~Churchland, JP~Cunningham, MT~Kaufman, JD~Foster, P~Nuyujukian, SI~Ryu, and
  KV~Shenoy.
\newblock Neural population dynamics during reaching.
\newblock {\em Nature}, 487(7405):51--56, 2012.

\bibitem{SVOBODA201833}
Karel Svoboda and Nuo Li.
\newblock Neural mechanisms of movement planning: motor cortex and beyond.
\newblock {\em Current Opinion in Neurobiology}, 49:33--41, 2018.
\newblock Neurobiology of Behavior.

\bibitem{lai2020acute}
G.~Lai, JP. Langevin, RJ. Koek, SE. Krahl, AA. Bari, and JWY. Chen.
\newblock Acute effects and the dreamy state evoked by deep brain electrical
  stimulation of the amygdala: Associations of the amygdala in human dreaming,
  consciousness, emotions, and creativity.
\newblock {\em Frontiers in Human Neuroscience}, 14:61, Feb 2020.

\bibitem{collee2023localization}
E~Coll{\'e}e, A~Vincent, E~Visch-Brink, E~De~Witte, C~Dirven, and D~Satoer.
\newblock Localization patterns of speech and language errors during awake
  brain surgery: a systematic review.
\newblock {\em Neurosurgical Review}, 46(1):38, Jan 2023.

\bibitem{ng2021disrupting}
S.~Ng, G.~Herbet, and AL. et~al. Lemaitre.
\newblock Disrupting self-evaluative processing with electrostimulation mapping
  during awake brain surgery.
\newblock {\em Scientific Reports}, 11:9386, 2021.

\bibitem{fried1991functional}
Itzhak Fried, Amiram Katz, Gregory McCarthy, Karen~J Sass, Philip Williamson,
  Susan~S Spencer, and Dennis~D Spencer.
\newblock Functional organization of human supplementary motor cortex studied
  by electrical stimulation.
\newblock {\em Journal of Neuroscience}, 11(11):3656--3666, Nov 1991.

\bibitem{deisseroth2015optogenetics}
Karl Deisseroth.
\newblock Optogenetics: 10 years of microbial opsins in neuroscience.
\newblock {\em Nature Neuroscience}, 18(8):1213--1225, 2015.

\bibitem{marshel2019cortical}
James~H Marshel, Yongsoo~S Kim, Timothy~A Machado, Sean Quirin, Brendan Benson,
  Jonathan Kadmon, Chandramouli Raja, Artur Chibukhchyan, Charu Ramakrishnan,
  Masafumi Inoue, Justin~C Shane, Daniel~J McKnight, Shin Yoshizawa, Hideaki~E
  Kato, Surya Ganguli, and Karl Deisseroth.
\newblock Cortical layer-specific critical dynamics triggering perception.
\newblock {\em Science}, 365(6453):eaaw5202, Aug 2019.

\bibitem{CARRILLOREID2019447}
Luis Carrillo-Reid, Shuting Han, Weijian Yang, Alejandro Akrouh, and Rafael
  Yuste.
\newblock Controlling visually guided behavior by holographic recalling of
  cortical ensembles.
\newblock {\em Cell}, 178(2):447--457.e5, 2019.

\bibitem{ramirez2015activating}
Steve Ramirez, Xiaojing Liu, Christopher MacDonald, et~al.
\newblock Activating positive memory engrams suppresses depression-like
  behaviour.
\newblock {\em Nature}, 522:335--339, 2015.

\bibitem{seth2022theories}
Anil~K Seth and Tim Bayne.
\newblock Theories of consciousness.
\newblock {\em Nature Reviews Neuroscience}, 23:439--452, 2022.

\bibitem{Mainen1995}
Zachary~F. Mainen and Terrence~J. Sejnowski.
\newblock Reliability of spike timing in neocortical neurons.
\newblock {\em Science}, 268(5216):1503--1506, Jun 1995.

\bibitem{Ratcliff2001}
Roger Ratcliff.
\newblock Putting noise into neurophysiological models of simple decision
  making.
\newblock {\em Nature Neuroscience}, 4(4):336, 2001.

\bibitem{Ratcliff2008}
Roger Ratcliff and Gail McKoon.
\newblock The diffusion decision model: theory and data for two-choice decision
  tasks.
\newblock {\em Neural Computation}, 20(4):873--922, Apr 2008.

\bibitem{tong2023single}
Ling Tong, Shuo Han, Yusheng Xue, Mengjiao Chen, Fei Chen, Wei Ke, Yousheng
  Shu, Nan Ding, Joerg Bewersdorf, Zhong-Wei Zhou, Peng Yuan, and Jaime
  Grutzendler.
\newblock Single cell in vivo optogenetic stimulation by two-photon excitation
  fluorescence transfer.
\newblock {\em iScience}, 26(10):107857, Sep 2023.

\bibitem{northoff2020neural}
Georg Northoff and Victor Lamme.
\newblock Neural signs and mechanisms of consciousness: is there a potential
  convergence of theories of consciousness in sight?
\newblock {\em Neuroscience \& Biobehavioral Reviews}, 118:568--587, 2020.

\bibitem{kent2019duration}
L.~Kent.
\newblock Duration perception versus perception duration: a proposed model for
  the consciously experienced moment.
\newblock {\em Timing \& Time Perception}, 7:1--14, 2019.

\bibitem{Fair1979-FAICAT}
David Fair.
\newblock Causation and the flow of energy.
\newblock {\em Erkenntnis}, 14(3):219--250, 1979.

\bibitem{Salmon1984-SALSEA}
Wesley~C. Salmon.
\newblock {\em Scientific Explanation and the Causal Structure of the World}.
\newblock Princeton University Press, 1984.

\bibitem{Dowe2000-DOWPC-2}
Phil Dowe.
\newblock {\em Physical Causation}.
\newblock Cambridge University Press, New York, 2000.

\bibitem{fazekas2021dynamical}
Peter Fazekas, Balazs Gyenis, Gábor Hofer-Szabó, and Gergely Kertesz.
\newblock A dynamical systems approach to causation.
\newblock {\em Synthese}, 198(11):6065--6087, 2021.

\bibitem{pouget2003inference}
Alexandre Pouget, Peter Dayan, and Richard~S Zemel.
\newblock Inference and computation with population codes.
\newblock {\em Annual Review of Neuroscience}, 26:381--410, 2003.
\newblock Epub 2003 Apr 10.

\bibitem{averbeck2006neural}
Bruno~B Averbeck, Peter~E Latham, and Alexandre Pouget.
\newblock Neural correlations, population coding and computation.
\newblock {\em Nature Reviews Neuroscience}, 7(5):358--366, May 2006.

\bibitem{vyas2020computation}
Surya Vyas, Matthew~D Golub, David Sussillo, and Krishna~V Shenoy.
\newblock Computation through neural population dynamics.
\newblock {\em Annual Review of Neuroscience}, 43:249--275, Jul 2020.

\bibitem{mazor2005transient}
Ofer Mazor and Gilles Laurent.
\newblock Transient dynamics versus fixed points in odor representations by
  locust antennal lobe projection neurons.
\newblock {\em Neuron}, 48(4):661--673, Nov 2005.

\bibitem{gardner2022toroidal}
Russell~J Gardner, Eilif Hermansen, Marius Pachitariu, et~al.
\newblock Toroidal topology of population activity in grid cells.
\newblock {\em Nature}, 602:123--128, 2022.

\bibitem{humphries2020strong}
Mark~D Humphries.
\newblock Strong and weak principles of neural dimension reduction.
\newblock 2020.

\bibitem{Kitcher1988}
Patricia Kitcher.
\newblock Marr's computational theory of vision.
\newblock {\em Philosophy of Science}, 55(1):1--24, 1988.

\bibitem{HORNIK1989359}
Kurt Hornik, Maxwell Stinchcombe, and Halbert White.
\newblock Multilayer feedforward networks are universal approximators.
\newblock {\em Neural Networks}, 2(5):359--366, 1989.

\bibitem{csaji2001approximation}
Balázs~Csanád Csáji.
\newblock Approximation with artificial neural networks.
\newblock Master's thesis, Eötvös Loránd University, Faculty of Sciences,
  Hungary, 2001.
\newblock MSc thesis.

\bibitem{Chalmers2010-CHATSA}
David~J. Chalmers.
\newblock The singularity: A philosophical analysis.
\newblock {\em Journal of Consciousness Studies}, 17(9-10):9--10, 2010.

\bibitem{GOERTZEL2007}
Ben Goertzel.
\newblock Human-level artificial general intelligence and the possibility of a
  technological singularity: A reaction to ray kurzweil's the singularity is
  near, and mcdermott's critique of kurzweil.
\newblock {\em Artificial Intelligence}, 171(18):1161--1173, 2007.
\newblock Special Review Issue.

\bibitem{Goertzel2012}
Ben Goertzel and Matthew Ikle'.
\newblock Mind uploading, introduction.
\newblock {\em International Journal of Machine Consciousness}, 04(01):1--3,
  2012.

\bibitem{Makin2019}
S~Makin.
\newblock The four biggest challenges in brain simulation.
\newblock {\em Nature Outlook}, July 2019.

\bibitem{Bamford2012-BAMAFF}
Sim Bamford.
\newblock A framework for approaches to transfer of a mind's substrate.
\newblock {\em International Journal of Machine Consciousness}, 4(01):23--34,
  2012.

\bibitem{Haggstrom2017}
Olle Häggström.
\newblock Aspects of mind uploading.
\newblock {\em Proceedings}, 1(3), 2017.

\bibitem{Pigliucci2014}
Massimo Pigliucci.
\newblock {\em Mind Uploading: A Philosophical Counter-Analysis}, chapter~7,
  pages 119--130.
\newblock John Wiley and Sons, Ltd, 2014.

\bibitem{Moor1988}
J.H. Moor.
\newblock Testing robots for qualia.
\newblock In H.R. Otto and J.A. Tuedio, editors, {\em Perspectives on Mind}. D.
  Reidel Publishing Company, Dordrecht/ Boston/ Lancaster/ Tokyo, 1988.

\bibitem{Prinz2003}
Jesse~J. Prinz.
\newblock Level-headed mysterianism and artificial experience.
\newblock In Owen Holland, editor, {\em Machine Consciousness}. Imprint
  Academic, Exeter, 2003.

\bibitem{Searle1980}
John~R. Searle.
\newblock Minds, brains and programs.
\newblock {\em Behavioral and Brain Sciences}, 3:417--457, 1980.

\end{thebibliography}
\bibliographystyle{unsrt}

}

\newpage
\appendix
\section*{Appendix / supplemental material}
\section{Terminology}\label{sec:terminology}
Below we provide descriptions and working definitions of some of the terms in our work.

\textit{Sufficient accurate emulator or model}. Sufficiently accurate is an empirically derivable term. It refers to whether a model behaves (or generates output) that is so close to our biological system that we cannot distinguish the differences. As we point out in our main text, a central implication of ET is that we will need increasingly nuanced tests for behavior and conscious states and, more critically, we may need to set these thresholds empirically - i.e. without a clear theory. 

\textit{Perfect models}. Similar to "sufficiently accurate", we reserve this term to mean a synthetic or computational model that is indistinguishable from the organism or neural system it is seeking to model. 

\textit{Sufficiently large or big datasets}. This term refers to the dataset size required to generate a sufficiently accurate emulator.  As explained in the main text, ET proposes that even fMRI data might be good enough to generate accurate emulators, but it is likely that significantly more data (and perhaps lower noise fMRI datasets) will be required. This would be in contrast to using data from single neuronal activations for emulators which in principle contain more information and less noise and could be smaller in size.  

\textit{Phenomenal or conscious states}. We use this term to refer to the internal, first-person, observation that is generally not accessible from the third person perspective. We note that did not distinguish between phenomenal and access consciousness and that we referred to phenomenal states as a type of "reportable" state.  While some have argued that such states may be in principle experienced - but not reportable, it is central to our work which only requires some way to measure conscious states - usually from a third person perspective. 

\textit{Coarse grained or exogenous activation}. These terms refer to the lack of precision in  direct-electrical-stimulation or optogenetic activation. Such exogenous neural activations generally target many neurons and neural circuits. Even in cases where a single neuron is activated, such inputs are still artificial - relative to an endogenous activation of the neuron which generally involves likely dozens but possibly hundreds of inputs form anterograde neurons and increasingly more from those cells' parental inputs.

\textit{Neural state}. This term refers usually to the simultaneous activity of the entire neural system we study.  In systems neuroscience, a neural state is also called a population vector or state and can represent the activity of each neuron recorded (e.g. being active or not) within a 1millisecond or larger time windows (depending on analysis carried out). 

\textit{Behaviors or actions}. We used this term to refer to the behavior of an animals, such as a mouse moving its body, or its limbs in space. 

\textit{Emulator}. Predictive (joint) models of animal behavior and whole-organism neural dynamics generated from different spatio-temporal scaled data. We generally intend the term emulator to refer to the model and its physical instantiation (i.e. operation) on a physical computer system. 

\textit{Emulator theory}. A theoretical framework for measuring and replicating the capacities of biological organisms solely via predictive models that do not require explicit mechanisms of how neural (sub)systems interact.

\section{The physics of neural causality}\label{sec:physics_of_causality}
While ET may not be required to explain how to generate consciousness - it may require to describe how neural causality works in biophysical organisms (e.g. mammals) and whether such causality can be present in artificial ones. Neuroscience does not directly adress the philosophy or physics of neural state causality - and only provides high-level candidate theories of causality (nearly all correlation-based) (e.g. \cite{seth2022theories}). Arguably, there is an implicit reductionist assumption to all these theories - that consciousness is caused by specific pathways and it emerges from the activity of lower-level or more fundamental systems. 

Here we seek to address the question of what causality may look like in "carbon-based" organisms and whether our best physics or philosophical theories of causality preclude artificial organisms from similar causal phenomena. We begin by briefly commenting on the science and physics of causality - a field without a specific consensus especially with respect to complex physical systems.  We then discuss causality in emulations and the sufficient conditions for recreation of all source capacities in an emulated system.  

\subsection{How are effects "caused" in complex dynamical (nervous) systems?}
The philosophy and physics of causality is still a developing field with the most common proposals for what "causality" means focusing on the transference or transmission of physical properties from one part to another or conservation of quantities across interactions \cite{Fair1979-FAICAT,Salmon1984-SALSEA, Dowe2000-DOWPC-2}. In these frameworks, we can conceptualize of a neuron as \textit{causing} another neuron to spike by passing on some physical properties (e.g. sufficient volume of glutamate molecules, or sufficiently large membrane current). However, many higher-order capacities (e.g. perception or cognition) simply cannot be reduced to cause-effect explanations between single neurons due to the infeasibility of tracking all the physics - but also the fact that cumulative effects of many neurons lead to an emergent property (i.e. cognition) that is dynamical and almost always independent of any individual single neuron. Some conceptual proposals for "causality by dynamics" have been put forward \cite{fazekas2021dynamical}. In these dynamical systems explanations, "[c]ause and effect states are ... regions of the state space, and the causal claim asserts a connection between these two regions". More specifically, "what makes a causal claim true is how the physical states it picks out are related by time evolution...". In simpler terms, in a complex dynamical system an \textit{effect} is caused if the time evolution of the \textit{causal} state is likely or "sufficiently" likely to lead to the effect state. Importantly, this definition is permissive enough to bridge the biophysical-cognitive divide.

Computational neuroscientists have already identified how the dynamics of populations of neurons are at least supportive of, if not central to, brain function \cite{pouget2003inference, averbeck2006neural, vyas2020computation}. In this "computation by population" paradigm “a neural population constitutes a dynamical system that, through its temporal evolution, performs a computation to, for example, generate and control movement". There have been several successful models using population level dynamical trajectories (e.g. \cite{mazor2005transient, churchland2012neural, gardner2022toroidal}).\footnote{For example \cite{mazor2005transient}, when presented with different odors, the \textit{individual} principal neurons (PNs) of the locust’s olfactory system exhibit a brief transient ON signal that is quite similar across all neurons and odors. However, when pooling the neurons together into a “population vector”, principal-component-analysis (PCA) visualizations show that odor processing has a substantially different dynamical trajectory at the population level for the different odors including fixed points and within-odor trial variance well below the inter-odor variance.}. And while population dynamics are usually visualized using dimensionality reduction tools such as PCA, there are suggestions that such tools are more than interpretation aids and “show us how neural circuits actually operate and compute” \cite{humphries2020strong}. This implies that in some sense computation is achieved by populations - not just individual neurons. 

Given this we draw two conclusions. First, we lack a complete theory of physical causality especially when it comes to complex dynamical systems. Second, and more central to our question, in a complex dynamical system such as the brain, the (i) physical states and (ii) their dynamical evolution are the only things we can measure and they \textit{must account} for the emergence of macro-states such as consciousness from micro-states. 

\subsection{Can software systems (emulators or simulators) cause or experience cognition?}
The above discussion suggests that \textit{if} we can recreate both neural states and their dynamical changes in a model or a "simulation" - we are essentially recreating all the components of the original systems and there is nothing else to explain about such systems. While such simulations could solve the same tasks of the source systems - would such simulations "experience" cognitive states? 

\begin{figure}
  \centering
  \fbox{\rule[-.25cm]{0cm}{0cm} \rule[-.25cm]{0cm}{0cm}
  \includegraphics[width=.8\linewidth]{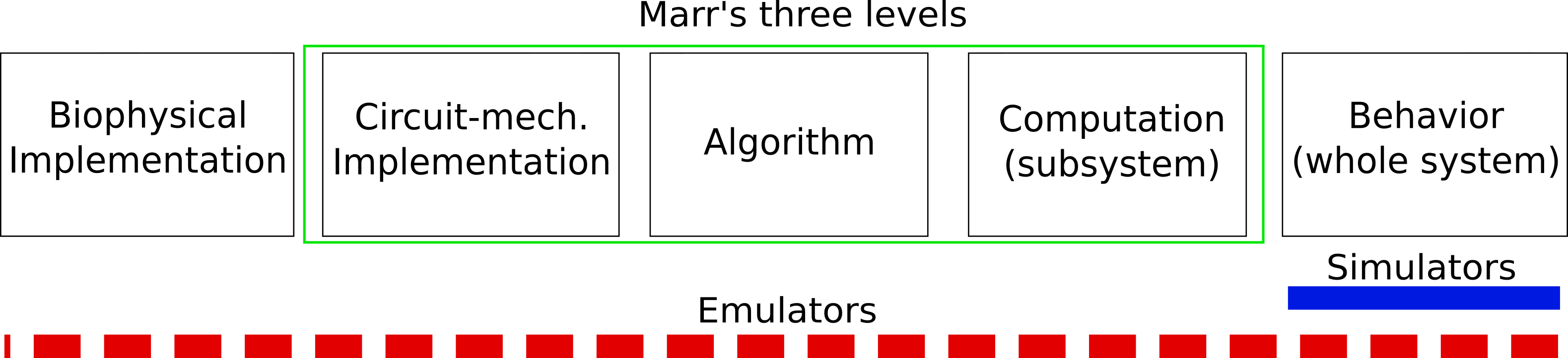}}
  
  \caption{\textbf{Different depths of models.} Relative to David Marr's \cite{marr1982vision} three levels of modeling (green rectangle), emulators model all levels of a biophysical system, while simulators model only behavior.}  
  \label{fig:depth_models}
   
\end{figure}

In our view, there is nothing in physics prohibiting complete emulation or simulation  including internal states of consciousness of self-organizing-systems. We can formulate this claim around David Marr's  well known computational theory of vision \cite{marr1982vision} containing three levels: computational, algorithmic and implementational (Fig \ref{fig:depth_models}). The distinctions are that "[t]he Computational theory tells us what function is being computed, for example, the square root function; the algorithmic level provides a means of carrying out the computation, for example, Newton's method of approximating square roots" whereas at "the implementation level... the same algorithm can be implemented in different hardware" \cite{Kitcher1988}.  In this space, an emulator is a computational system that captures both the input-output functions of a source-system and possibly its algorithmic, circuit-mechanistic and biophysical-substrate implementations (Fig 4).  Thus emulators seek a - potentially - biophysical copy of an original source system, while simulators are largely interested in matching the computation level only (i.e. input-output function) of the system studied. 

We propose this as a working definition, amenable to change, of how to distinguish between software vs. hardware-software models of biophysical systems. Our central claim, however, is that in the infinite - or "sufficiently granular" - limit emulators will contain all the components of the biophysical systems they are seeking to model and must necessarily also have all their capacities including internal conscious states. More specific to this section, \textit{emulators of dynamical systems cannot be distinguished from the dynamical systems themselves}. The central weakness or unknown, however, that we do not know at what level of brain emulation we get consciousness.

\subsection{Conclusion: rethinking the need for infinite reductionism in the causality of non-physical states}
In this section we highlighted the limited conceptual agreement on the physics of causality and the implications on limitations for emulating complex biophysical organisms. Our framework suggests that a biophysical emulator could generate indistinguishable neural but also behavioral dynamics from a source, e.g. optimal motor responses and planning. 





\section{Ideal emulators}\label{sec:ideal_emulators}

In the main text we proposed \textit{ideal emulators} as systems that - in the infinite granularity and infinite data-size limit - can perfectly model all the necessary and sufficient components of a biological organism that generate behavior and conscious states.\footnote{We note that we generally mean ideal \textit{neural} emulators - but drop the term "neural" for simplicity.} As an example, an ideal emulator can be thought of as a complete biophysical copy of an animal's nervous system, from atoms to synapses to neuron connections - that can be reconstructed based on access to "big" data. However, as we are interested in predicting \textit{behavior} and \textit{conscious states} of biological organisms - not synaptic, molecular or atomic interactions - an ideal emulator will focus on learning to model these two types of output primarily. The existence of ideal emulators is essentially a restatement of the universal approximation theorem\cite{HORNIK1989359, csaji2001approximation} but linking neural states to behavioral and phenomenal states. Within such a framework, the central axiom of ET is:





\begin{enumerate}[label={A\arabic*:}]
    \item \textbf{Ideal emulator are equivalent to their sources}. There is no empirical test that can distinguish between the behavior output or conscious states experienced by a biological organism and those generated by an ideal emulator of such an organism. 
\end{enumerate}

We first note that $A1$ is a statement about how we \textit{measure} conscious states - in an emulator or a biological organism - rather than how conscious states \textit{emerge} in a biological organism. That is, $A1$ states that consciousness must be present (by definition) in an ideal emulator which is trained to predict the behavior and conscious states of a biological organism - even if those are never explicitly represented or even known by a human model designer. 

Within the "levels of analysis" framework provided by David Marr \cite{marr1982vision} 
an ideal emulator is a system that \textit{can} model all components of an organisms including the subsystem (e.g. vision) level computations, their algorithms, and circuit-level and even biophysical implementations (such as intracellular mechanisms and dynamics)(Fig \ref{fig:depth_models}). 

In contrast, we define a \textit{simulator} as a system that models the behavioral outputs of a biological system without any specific "lower-level" constraints.\footnote{We provide a discussion on zombie consciousness in the appendix}
\section{MSD: model system divergence hypothesis}\label{sec:MSD_model_indistinguish}

\begin{figure}
  \centering
  \fbox{\rule[-.25cm]{0cm}{0cm}   \rule[-.25cm]{0cm}{0cm}
  \includegraphics[width=.95\linewidth]{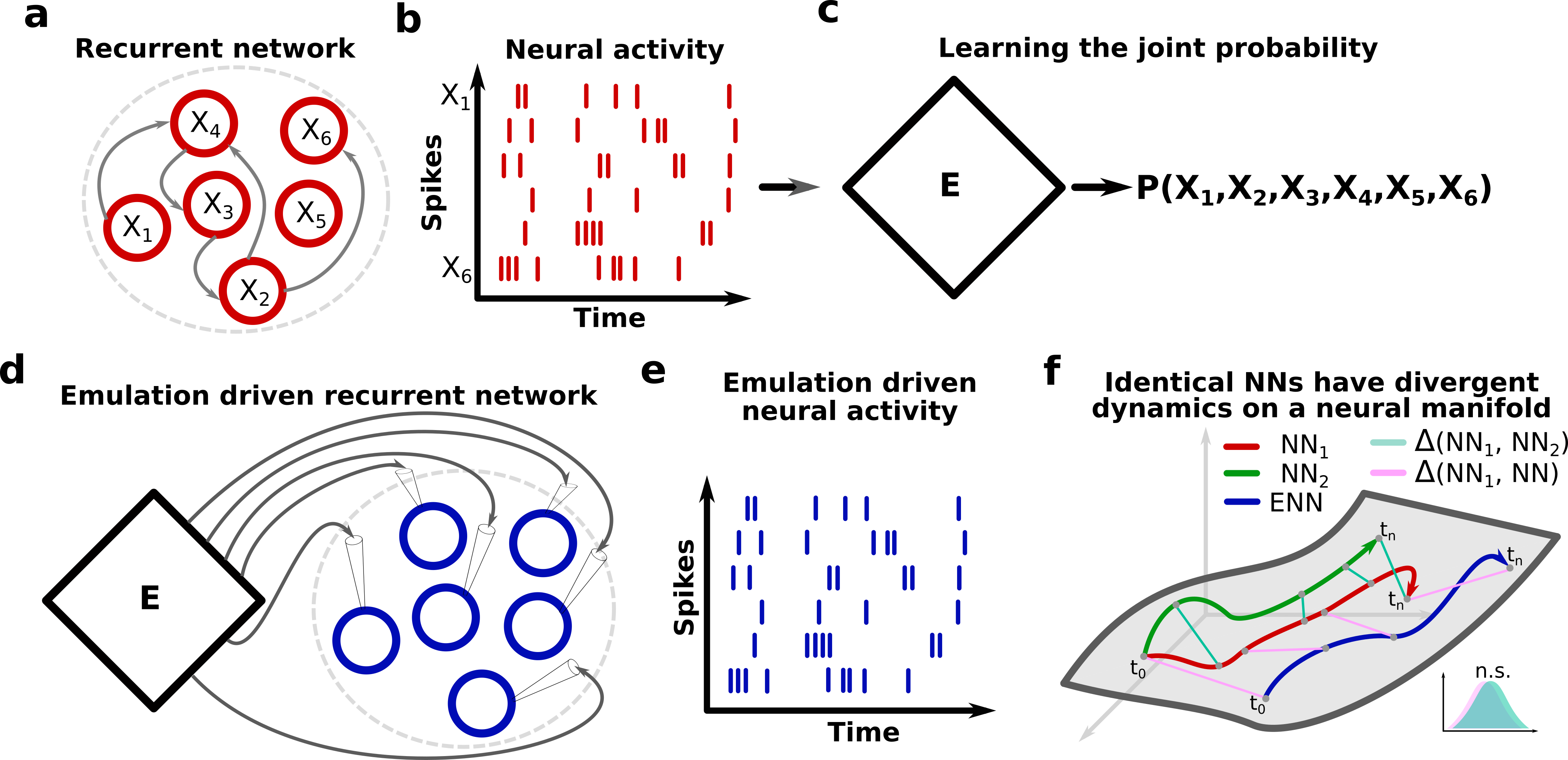}}
 
  \caption{\textbf{Emulators learn joint distributions and activate neuronal populations}. \textbf{(a)}. Six-node recurrent neural network. \textbf{(b)}. Proposed spike trains for network in (a). \textbf{(c)}. Emulators learn the joint probability distribution from empirical observations. \textbf{(d)}. Emulators can now drive the neural activity in a similar neural network that has no endogenous connections. \textbf{e}. Proposed spike trains obtained from (d) are similar to those in the endogenous case in (b). \textbf{f}. The evolution of two identical dynamical systems driven ground-truth mechanistic-based models (NN$_1$ and NN$_2$ is as divergent (or different) as the evolution of an emulator based dynamical system (ENN) from the two models (inset: putative distributions of path differences over time).}
  \label{fig:dynamical_manifold}

\end{figure}

The model system divergence hypothesis (MSD) is essentially a statement about the inherent noise and chaotic dynamics of the neural systems we seek to model. 

Given a neural network (e.g. an RNN) (Fig \ref{fig:dynamical_manifold}a) that generates spiking time series (Fig \ref{fig:dynamical_manifold}b) - a neural emulator is a model that learns the joint probability of the neural time series (i.e. spike rasters here) and additionally behavior outputs\footnote{We note that organism-level behavior outputs is in principle encoded in the neural states of the organism, and this secondary requirement is somewhat redundant.} (Fig \ref{fig:dynamical_manifold}c). As argued in the main text, such a model can artificially drive the origial biological network (Fig \ref{fig:dynamical_manifold}d) to generate similar (or identical) spiking time series (Fig \ref{fig:dynamical_manifold}e). 
Thus, given these inherent components of biological neural networks, two systems driven by NNs with identical connectivity weights and starting states that evolve on a constrained manifold - can nonetheless diverge substantially (Fig \ref{fig:dynamical_manifold}f - green and red curves). MSD states that the it is not possible to statistically distinguish (by empirical measurements alone) between a dynamical system built on ground-truth mechanistic models of biological system and a (sufficiently accurate) emulator-driven NNs (Fig \ref{fig:dynamical_manifold}f; inset histograms).

MSD thus  specifically highlights that relying on "precise" or accurate dynamics of neural systems models may not be enough to distinguish ground-truth mechanistic models of nervous systems and increasingly accurate emulators (or any other class of models).  In particular, as emulators become increasingly precise in capturing the dynamics of a biological organism - there may be no clear or discrete moment when complex capacities such as "conscious" states are being generated.



\section{Relation to previous work}\label{sec:previous_work}
Given the various interdisciplinary notions discussed in our work, we selected to focus mostly (and only at a high level) on the literature dealing with the philosophy and phenomenology of conscious states and topics related to machine consciousness.  

A central theme in artificial intelligence that is linked to understanding brain function is whether we can copy, emulate, or upload brains or minds to digital systems. In this field there increase discussions around humans minds being copied to machines \cite{Chalmers2010-CHATSA} or "mind-uploading" \cite{GOERTZEL2007, Goertzel2012}. Even setting aside the limited scientific progress on this front - there are  significant and unresolved conceptual challenges \cite{Makin2019} including lack of frameworks \cite{Bamford2012-BAMAFF}, the lack of clarity around whether digital copies will experience consciousness or have a personal identity \cite{Haggstrom2017, Pigliucci2014}. In parallel,while we still don't know the necessary or sufficient conditions for conscious states in biological organisms, there are suggestions of a scattershot approach \cite{butlin2023consciousness} to increase the probability of this. 

Progress in machine consciousness has also been framed around core principles \cite{GAMEZ2008} including creating machines whose behavior and cognitive characteristics are associated with consciousness, have similar architectures to conscious systems (i.e. humans) or actually experience phenomenal conscious states. For the later, the believe is that the "hard" problem of consciousness needs to be solved first \cite{Chalmers2010-CHATSA}. Others like \cite{Moor1988} and \cite{Prinz2003} suggest that it is not possible to know whether artificial systems can be conscious because we can not isolate or separate the factors for consciousness. In contrast, ET proposes  that we do not have to solve hard problem to create conscious machines nor have similar architectures as long the the machines outward behavior matches that of the source organisms - and we achieve such behaviors via modeling neural dynamics. 

Lastly, \cite{Searle1980} in the famous "Chinese room" experiment suggests that an artificial system that does "mindless" translation by rules (or lookup tables) does not amount to understand a language, for example, and therefore cannot have phenomenal states. ET propose that if the artificial system produces the "mindless" translation by modeling the causal components of a real - e.g. biological organism that can translate - then there is an equivalence between the biological and the artificial system and it is difficult to distinguish between the two systems - aside from some molecular components of their makeup.





\newpage
\section*{NeurIPS Paper Checklist}


\begin{enumerate}
\item {\bf Claims}
    \item[] Question: Do the main claims made in the abstract and introduction accurately reflect the paper's contributions and scope?
    \item[] Answer: \answerYes{} 
    \item[] Justification: the Abstract and Introduction frame the contribution of the paper relative to existing research. We clearly define the scope and contribution the introduction and a brief discussion of previous work (with a broader discussion in the Appendix). 
    \item[] Guidelines:
    \begin{itemize}
        \item The answer NA means that the abstract and introduction do not include the claims made in the paper.
        \item The abstract and/or introduction should clearly state the claims made, including the contributions made in the paper and important assumptions and limitations. A No or NA answer to this question will not be perceived well by the reviewers. 
        \item The claims made should match theoretical and experimental results, and reflect how much the results can be expected to generalize to other settings. 
        \item It is fine to include aspirational goals as motivation as long as it is clear that these goals are not attained by the paper. 
    \end{itemize}

\item {\bf Limitations}
    \item[] Question: Does the paper discuss the limitations of the work performed by the authors?
    \item[] Answer: \answerYes{} 
    \item[] Justification: The paper discusses implications and limitations of our work in a separate section. As our work is conceptual and theoretical in nature, we provide theoretical limitations and conceptual challenges in building the types of systems and models we propose. 
    \item[] Guidelines:
    \begin{itemize}
        \item The answer NA means that the paper has no limitation while the answer No means that the paper has limitations, but those are not discussed in the paper. 
        \item The authors are encouraged to create a separate "Limitations" section in their paper.
        \item The paper should point out any strong assumptions and how robust the results are to violations of these assumptions (e.g., independence assumptions, noiseless settings, model well-specification, asymptotic approximations only holding locally). The authors should reflect on how these assumptions might be violated in practice and what the implications would be.
        \item The authors should reflect on the scope of the claims made, e.g., if the approach was only tested on a few datasets or with a few runs. In general, empirical results often depend on implicit assumptions, which should be articulated.
        \item The authors should reflect on the factors that influence the performance of the approach. For example, a facial recognition algorithm may perform poorly when image resolution is low or images are taken in low lighting. Or a speech-to-text system might not be used reliably to provide closed captions for online lectures because it fails to handle technical jargon.
        \item The authors should discuss the computational efficiency of the proposed algorithms and how they scale with dataset size.
        \item If applicable, the authors should discuss possible limitations of their approach to address problems of privacy and fairness.
        \item While the authors might fear that complete honesty about limitations might be used by reviewers as grounds for rejection, a worse outcome might be that reviewers discover limitations that aren't acknowledged in the paper. The authors should use their best judgment and recognize that individual actions in favor of transparency play an important role in developing norms that preserve the integrity of the community. Reviewers will be specifically instructed to not penalize honesty concerning limitations.
    \end{itemize}

\item {\bf Theory Assumptions and Proofs}
    \item[] Question: For each theoretical result, does the paper provide the full set of assumptions and a complete (and correct) proof?
    \item[] Answer: \answerYes{} 
    \item[] Justification: We provide our assumptions for our theoretical proposals in detail in each of the sections. In particular, emulators assume the ability to record specific types of neural dynamics and have sufficiently large datasets.  We have also collected some of these terms in an Appendix specific on terminology.
    \item[] Guidelines:
    \begin{itemize}
        \item The answer NA means that the paper does not include theoretical results. 
        \item All the theorems, formulas, and proofs in the paper should be numbered and cross-referenced.
        \item All assumptions should be clearly stated or referenced in the statement of any theorems.
        \item The proofs can either appear in the main paper or the supplemental material, but if they appear in the supplemental material, the authors are encouraged to provide a short proof sketch to provide intuition. 
        \item Inversely, any informal proof provided in the core of the paper should be complemented by formal proofs provided in appendix or supplemental material.
        \item Theorems and Lemmas that the proof relies upon should be properly referenced. 
    \end{itemize}

    \item {\bf Experimental Result Reproducibility}
    \item[] Question: Does the paper fully disclose all the information needed to reproduce the main experimental results of the paper to the extent that it affects the main claims and/or conclusions of the paper (regardless of whether the code and data are provided or not)?
    \item[] Answer: \answerNA{} 
    \item[] Justification: We did not run any specific experiments.
    \item[] Guidelines:
    \begin{itemize}
        \item The answer NA means that the paper does not include experiments.
        \item If the paper includes experiments, a No answer to this question will not be perceived well by the reviewers: Making the paper reproducible is important, regardless of whether the code and data are provided or not.
        \item If the contribution is a dataset and/or model, the authors should describe the steps taken to make their results reproducible or verifiable. 
        \item Depending on the contribution, reproducibility can be accomplished in various ways. For example, if the contribution is a novel architecture, describing the architecture fully might suffice, or if the contribution is a specific model and empirical evaluation, it may be necessary to either make it possible for others to replicate the model with the same dataset, or provide access to the model. In general. releasing code and data is often one good way to accomplish this, but reproducibility can also be provided via detailed instructions for how to replicate the results, access to a hosted model (e.g., in the case of a large language model), releasing of a model checkpoint, or other means that are appropriate to the research performed.
        \item While NeurIPS does not require releasing code, the conference does require all submissions to provide some reasonable avenue for reproducibility, which may depend on the nature of the contribution. For example
        \begin{enumerate}
            \item If the contribution is primarily a new algorithm, the paper should make it clear how to reproduce that algorithm.
            \item If the contribution is primarily a new model architecture, the paper should describe the architecture clearly and fully.
            \item If the contribution is a new model (e.g., a large language model), then there should either be a way to access this model for reproducing the results or a way to reproduce the model (e.g., with an open-source dataset or instructions for how to construct the dataset).
            \item We recognize that reproducibility may be tricky in some cases, in which case authors are welcome to describe the particular way they provide for reproducibility. In the case of closed-source models, it may be that access to the model is limited in some way (e.g., to registered users), but it should be possible for other researchers to have some path to reproducing or verifying the results.
        \end{enumerate}
    \end{itemize}

\item {\bf Open access to data and code}
    \item[] Question: Does the paper provide open access to the data and code, with sufficient instructions to faithfully reproduce the main experimental results, as described in supplemental material?
    \item[] Answer: \answerNA{} 
    \item[] Justification: We did not run any experiments at this time.
    \item[] Guidelines:
    \begin{itemize}
        \item The answer NA means that paper does not include experiments requiring code.
        \item Please see the NeurIPS code and data submission guidelines (\url{https://nips.cc/public/guides/CodeSubmissionPolicy}) for more details.
        \item While we encourage the release of code and data, we understand that this might not be possible, so “No” is an acceptable answer. Papers cannot be rejected simply for not including code, unless this is central to the contribution (e.g., for a new open-source benchmark).
        \item The instructions should contain the exact command and environment needed to run to reproduce the results. See the NeurIPS code and data submission guidelines (\url{https://nips.cc/public/guides/CodeSubmissionPolicy}) for more details.
        \item The authors should provide instructions on data access and preparation, including how to access the raw data, preprocessed data, intermediate data, and generated data, etc.
        \item The authors should provide scripts to reproduce all experimental results for the new proposed method and baselines. If only a subset of experiments are reproducible, they should state which ones are omitted from the script and why.
        \item At submission time, to preserve anonymity, the authors should release anonymized versions (if applicable).
        \item Providing as much information as possible in supplemental material (appended to the paper) is recommended, but including URLs to data and code is permitted.
    \end{itemize}

\item {\bf Experimental Setting/Details}
    \item[] Question: Does the paper specify all the training and test details (e.g., data splits, hyperparameters, how they were chosen, type of optimizer, etc.) necessary to understand the results?
    \item[] Answer: \answerNA{} 
    \item[] Justification: We did not run any experiments.
    \item[] Guidelines:
    \begin{itemize}
        \item The answer NA means that the paper does not include experiments.
        \item The experimental setting should be presented in the core of the paper to a level of detail that is necessary to appreciate the results and make sense of them.
        \item The full details can be provided either with the code, in appendix, or as supplemental material.
    \end{itemize}

\item {\bf Experiment Statistical Significance}
    \item[] Question: Does the paper report error bars suitably and correctly defined or other appropriate information about the statistical significance of the experiments?
    \item[] Answer: \answerNA{} 
    \item[] Justification: We did not run any experiments.
    \item[] Guidelines:
    \begin{itemize}
        \item The answer NA means that the paper does not include experiments.
        \item The authors should answer "Yes" if the results are accompanied by error bars, confidence intervals, or statistical significance tests, at least for the experiments that support the main claims of the paper.
        \item The factors of variability that the error bars are capturing should be clearly stated (for example, train/test split, initialization, random drawing of some parameter, or overall run with given experimental conditions).
        \item The method for calculating the error bars should be explained (closed form formula, call to a library function, bootstrap, etc.)
        \item The assumptions made should be given (e.g., Normally distributed errors).
        \item It should be clear whether the error bar is the standard deviation or the standard error of the mean.
        \item It is OK to report 1-sigma error bars, but one should state it. The authors should preferably report a 2-sigma error bar than state that they have a 96\% CI, if the hypothesis of Normality of errors is not verified.
        \item For asymmetric distributions, the authors should be careful not to show in tables or figures symmetric error bars that would yield results that are out of range (e.g. negative error rates).
        \item If error bars are reported in tables or plots, The authors should explain in the text how they were calculated and reference the corresponding figures or tables in the text.
    \end{itemize}

\item {\bf Experiments Compute Resources}
    \item[] Question: For each experiment, does the paper provide sufficient information on the computer resources (type of compute workers, memory, time of execution) needed to reproduce the experiments?
    \item[] Answer: \answerNA{} 
    \item[] Justification: We did not run any experiments.
    \item[] Guidelines:
    \begin{itemize}
        \item The answer NA means that the paper does not include experiments.
        \item The paper should indicate the type of compute workers CPU or GPU, internal cluster, or cloud provider, including relevant memory and storage.
        \item The paper should provide the amount of compute required for each of the individual experimental runs as well as estimate the total compute. 
        \item The paper should disclose whether the full research project required more compute than the experiments reported in the paper (e.g., preliminary or failed experiments that didn't make it into the paper). 
    \end{itemize}
    
\item {\bf Code Of Ethics}
    \item[] Question: Does the research conducted in the paper conform, in every respect, with the NeurIPS Code of Ethics \url{https://neurips.cc/public/EthicsGuidelines}?
    \item[] Answer: \answerYes{} 
    \item[] Justification: Our paper conforms with the ethics guidelines.
    \item[] Guidelines:
    \begin{itemize}
        \item The answer NA means that the authors have not reviewed the NeurIPS Code of Ethics.
        \item If the authors answer No, they should explain the special circumstances that require a deviation from the Code of Ethics.
        \item The authors should make sure to preserve anonymity (e.g., if there is a special consideration due to laws or regulations in their jurisdiction).
    \end{itemize}

\item {\bf Broader Impacts}
    \item[] Question: Does the paper discuss both potential positive societal impacts and negative societal impacts of the work performed?
    \item[] Answer: \answerNA{}
    \item[] Justification: Our work does not provide any immediate societal implications. 
    \item[] Guidelines:
    \begin{itemize}
        \item The answer NA means that there is no societal impact of the work performed.
        \item If the authors answer NA or No, they should explain why their work has no societal impact or why the paper does not address societal impact.
        \item Examples of negative societal impacts include potential malicious or unintended uses (e.g., disinformation, generating fake profiles, surveillance), fairness considerations (e.g., deployment of technologies that could make decisions that unfairly impact specific groups), privacy considerations, and security considerations.
        \item The conference expects that many papers will be foundational research and not tied to particular applications, let alone deployments. However, if there is a direct path to any negative applications, the authors should point it out. For example, it is legitimate to point out that an improvement in the quality of generative models could be used to generate deepfakes for disinformation. On the other hand, it is not needed to point out that a generic algorithm for optimizing neural networks could enable people to train models that generate Deepfakes faster.
        \item The authors should consider possible harms that could arise when the technology is being used as intended and functioning correctly, harms that could arise when the technology is being used as intended but gives incorrect results, and harms following from (intentional or unintentional) misuse of the technology.
        \item If there are negative societal impacts, the authors could also discuss possible mitigation strategies (e.g., gated release of models, providing defenses in addition to attacks, mechanisms for monitoring misuse, mechanisms to monitor how a system learns from feedback over time, improving the efficiency and accessibility of ML).
    \end{itemize}
    
\item {\bf Safeguards}
    \item[] Question: Does the paper describe safeguards that have been put in place for responsible release of data or models that have a high risk for misuse (e.g., pretrained language models, image generators, or scraped datasets)?
    \item[] Answer: \answerNA{} 
    \item[] Justification: Our work does not pose any immediate risks.
    \item[] Guidelines:
    \begin{itemize}
        \item The answer NA means that the paper poses no such risks.
        \item Released models that have a high risk for misuse or dual-use should be released with necessary safeguards to allow for controlled use of the model, for example by requiring that users adhere to usage guidelines or restrictions to access the model or implementing safety filters. 
        \item Datasets that have been scraped from the Internet could pose safety risks. The authors should describe how they avoided releasing unsafe images.
        \item We recognize that providing effective safeguards is challenging, and many papers do not require this, but we encourage authors to take this into account and make a best faith effort.
    \end{itemize}

\item {\bf Licenses for existing assets}
    \item[] Question: Are the creators or original owners of assets (e.g., code, data, models), used in the paper, properly credited and are the license and terms of use explicitly mentioned and properly respected?
    \item[] Answer: \answerNA{} 
    \item[] Justification: There are no applicable licenses required.
    \item[] Guidelines:
    \begin{itemize}
        \item The answer NA means that the paper does not use existing assets.
        \item The authors should cite the original paper that produced the code package or dataset.
        \item The authors should state which version of the asset is used and, if possible, include a URL.
        \item The name of the license (e.g., CC-BY 4.0) should be included for each asset.
        \item For scraped data from a particular source (e.g., website), the copyright and terms of service of that source should be provided.
        \item If assets are released, the license, copyright information, and terms of use in the package should be provided. For popular datasets, \url{paperswithcode.com/datasets} has curated licenses for some datasets. Their licensing guide can help determine the license of a dataset.
        \item For existing datasets that are re-packaged, both the original license and the license of the derived asset (if it has changed) should be provided.
        \item If this information is not available online, the authors are encouraged to reach out to the asset's creators.
    \end{itemize}

\item {\bf New Assets}
    \item[] Question: Are new assets introduced in the paper well documented and is the documentation provided alongside the assets?
    \item[] Answer: \answerNA{} 
    \item[] Justification: Our work does not introduce any new assets.
    \item[] Guidelines:
    \begin{itemize}
        \item The answer NA means that the paper does not release new assets.
        \item Researchers should communicate the details of the dataset/code/model as part of their submissions via structured templates. This includes details about training, license, limitations, etc. 
        \item The paper should discuss whether and how consent was obtained from people whose asset is used.
        \item At submission time, remember to anonymize your assets (if applicable). You can either create an anonymized URL or include an anonymized zip file.
    \end{itemize}

\item {\bf Crowdsourcing and Research with Human Subjects}
    \item[] Question: For crowdsourcing experiments and research with human subjects, does the paper include the full text of instructions given to participants and screenshots, if applicable, as well as details about compensation (if any)? 
    \item[] Answer: \answerNA{} 
    \item[] Justification: Our work did not involve crowdsourcing nor human subject research.
    \item[] Guidelines:
    \begin{itemize}
        \item The answer NA means that the paper does not involve crowdsourcing nor research with human subjects.
        \item Including this information in the supplemental material is fine, but if the main contribution of the paper involves human subjects, then as much detail as possible should be included in the main paper. 
        \item According to the NeurIPS Code of Ethics, workers involved in data collection, curation, or other labor should be paid at least the minimum wage in the country of the data collector. 
    \end{itemize}

\item {\bf Institutional Review Board (IRB) Approvals or Equivalent for Research with Human Subjects}
    \item[] Question: Does the paper describe potential risks incurred by study participants, whether such risks were disclosed to the subjects, and whether Institutional Review Board (IRB) approvals (or an equivalent approval/review based on the requirements of your country or institution) were obtained?
    \item[] Answer: \answerNA{} 
    \item[] Justification: Our work does not invovle crowdsourcing or human subject research
    \item[] Guidelines:
    \begin{itemize}
        \item The answer NA means that the paper does not involve crowdsourcing nor research with human subjects.
        \item Depending on the country in which research is conducted, IRB approval (or equivalent) may be required for any human subjects research. If you obtained IRB approval, you should clearly state this in the paper. 
        \item We recognize that the procedures for this may vary significantly between institutions and locations, and we expect authors to adhere to the NeurIPS Code of Ethics and the guidelines for their institution. 
        \item For initial submissions, do not include any information that would break anonymity (if applicable), such as the institution conducting the review.
    \end{itemize}

\end{enumerate}

\end{document}